# Evolution of Interlayer Coupling in Twisted MoS$_2$ Bilayers

Kaihui Liu[1,2]*, Liming Zhang[3]*, Ting Cao[1,4]*, Chenhao Jin[1], Diana Qiu[1,4], Qin Zhou[4], Alex Zettl[1,4,5], Peidong Yang[3,5,6], Steve G. Louie[1,4] and Feng Wang[1,4,5]

[1] Department of Physics, University of California at Berkeley, Berkeley, CA 94720, United States

[2] State Key Laboratory for Mesoscopic Physics and Collaborative Innovation Centre of Quantum Matter, Peking University, Beijing 100871, China

[3] Department of Chemistry, University of California at Berkeley, Berkeley, CA 94720, United States

[4] Materials Science Division, Lawrence Berkeley National Laboratory, Berkeley, California 94720, United States

[5] Kavli Energy NanoSciences Institute at the University of California, Berkeley and the Lawrence Berkeley National Laboratory, Berkeley, California, 94720, USA

[6] Department of Material Science and Engineering, University of California at Berkeley, Berkeley, CA 94720, United States

*These authors contribute equally to this work

Van der Waals (vdW) coupling is emerging as a powerful method to engineer and tailor physical properties of atomically thin two-dimensional (2D) materials. In graphene/graphene and graphene/boron-nitride structures it leads to interesting physical phenomena ranging from new van Hove singularities[1-4] and Fermi velocity renormalization[5,6] to unconventional quantum Hall effects[7] and Hofstadter's butterfly pattern[8-12]. 2D transition metal dichalcogenides (TMDCs), another system of predominantly vdW-coupled atomically thin layers[13,14], can also exhibit interesting but different coupling phenomena because TMDCs can be direct or indirect bandgap semiconductors[15,16]. Here, we present the first study on the evolution of interlayer coupling with twist angles in as-grown $MoS_2$ bilayers. We find that an indirect bandgap emerges in bilayers with any stacking configuration, but the bandgap size varies appreciably with the twist angle: it shows the largest redshift for AA- and AB-stacked bilayers, and a significantly smaller but constant redshift for all other twist angles. The vibration frequency of the out-of-plane phonon in $MoS_2$ shows similar twist angle dependence. Our observations, together with *ab initio* calculations, reveal that this evolution of interlayer coupling originates from the repulsive steric effects, which leads to different interlayer separations between the two $MoS_2$ layers in different stacking configurations.

Vertically stacked 2D atomic layers have attracted immense research interest recently. In graphene bilayers and graphene/boron nitride heterostructures, for example, the weak vdW interaction leads to new physical phenomena ranging from Fermi velocity renormalization to Hofstadter butterfly patterns[1-12]. In gapless graphene, the interlayer coupling depends sensitive on the interlayer twist angle, and the electronic properties around the K point is modified the most when the twist angle is small[5, 6]. However, the evolution of interlayer coupling in gapped 2D materials, such as the 2D transition metal dichalcogenides, can be very different. In this letter, we employ as-grown $MoS_2$ bilayers as a model system to investigate the evolution of interlayer electronic and mechanical coupling with the twist angle in 2D TMDC semiconductors. Monolayer $MoS_2$ exhibits many intriguing physical properties, including a direct optical band gap of ~1.8 eV in the visible range[15, 16], a strong exciton binding energy[17], and valley selective circular dichroism[18-20], which make it a promising material for next generation nano-electronics[21], photonics[22], photovoltaics[23], and valleytronics[18-20]. Previous studies on naturally AB-stacked $MoS_2$ bilayers show that the weak interlayer couplings can significantly modify the physical properties[24, 25], but little is known on how they evolve with the twist angle in arbitrarily stacked $MoS_2$ bilayers.

In our experiment, we grow triangle-shaped $MoS_2$ monolayers and bilayers with different stacking orders by chemical vapor deposition (CVD). Our growth is based on the recent progress in $MoS_2$ monolayer synthesis[26-28] but with modified growth conditions (See Methods). Our key control is to decrease the nucleation rate at the initial stage of growth to make the vertical layer-by-layer growth mode preferable. We can grow $MoS_2$ bilayers on various substrates, including mica, fused silica and $SiO_2/Si$ with yield as high as 30%. (The other 70% are mainly monolayers). Fig. 1e-h show, respectively, optical reflection images of a monolayer and bilayers with twist angles of $\theta = 0°, 15°, 60°$ on fused silica substrates. In bilayer samples, the smaller top layer is above the larger bottom layer, leading to a doubled optical contrast. Reflection spectra (Fig. 1i) of the samples show prominent resonances at 1.85 and 2.05 eV, characteristic of the direct exciton transitions in $MoS_2$ layers.

The twist angle of $MoS_2$ bilayer can be readily determined through the relative orientation of the top and bottom triangle shapes. In $MoS_2$ each Mo atom is covalently bonded to six S atoms in a trigonal prismatic arrangement (Fig. 1a), and the triangle shaped layers have been shown to terminate with Mo at zigzag edges in scanning tunneling microscopy (STM)[29], transmission electron microscopy (TEM)[28], and *ab initio* theoretical studies[30]. Therefore the orientation of each triangle is directly correlated with the microscopic crystal orientation of the $MoS_2$ layer, and we can determine the twist angle of a $MoS_2$ bilayer from the relative rotation of the two vertically stacked triangles (Fig 1b-d). We further confirmed the AA-stacking order in $\theta = 0°$ bilayers and the AB-stacking order in $\theta = 60°$ bilayers using second-harmonic generation (SHG) measurements (Fig. 1j), where SHG signal is strongly suppressed for AB-stacked bilayer due to its overall inversion symmetry[31](See supplementary S1 for more details).

The interlayer interaction can significantly modify both the electronic band structure and phonon vibrations in $MoS_2$ bilayers. We first investigate the evolution of the electronic coupling in different twisted $MoS_2$ bilayers using photoluminescence spectroscopy (PL). Fig. 2a shows the typical PL spectra of a $MoS_2$ monolayer and bilayers with twist angles of $\theta = 0°, 15°, 60°$. The monolayer PL spectrum features a prominent peak around 1.85 eV (peak I), which corresponds to the exciton transition at the K valley of the Brillouin Zone (Fig. 2c). For all bilayers, the direct exciton transition at the K valley of the original monolayer Brillouin zone (peak I) remains, but its PL intensity decreases significantly (to about 1/50 of the monolayer PL). In addition, a new peak at lower energy (peak II) appears. The emergence of this new peak II corresponds to an indirect bandgap resulting from the interlayer electronic coupling, in which the doubly degenerate valence band splits into two branches near the $\Gamma$ point and the upper branch rises to an energy higher than the valence band at the K point in the monolayer Brillouin zone (Fig. 2d). Therefore, the indirect bandgap (peak II) energy reflects directly the interlayer electronic coupling strength: the lower the indirect bandgap, the stronger the coupling strength. To investigate the evolution of electronic coupling in twisted bilayers in more detail, we systematically studied 44 $MoS_2$ bilayers with various twist angles. Figure 2b displays the transition energies of peak I and II in all measured bilayers, which exhibit interesting dependence on the twist angle. In contrast to a virtually unchanged peak I energy (direct exciton transition), the peak II energy (indirect transition) is lowest for perfect registered (i.e., AA- and AB-stacked) bilayers, and has a higher but nearly constant value for all other twist angles. This means that the interlayer electronic coupling is significant for all twist angles, but is the strongest in AA- or AB-stacked $MoS_2$ bilayers.

In addition to the electronic bandgaps, we investigated the evolution of phonon vibrations in twisted MoS$_2$ bilayers using Raman scattering spectroscopy. Fig. 3a shows the typical Raman spectra of a MoS$_2$ monolayer and bilayers with twist angles of $\theta = 0°, 15°, 60°$. Two prominent peaks are observed in the range of 360~440 cm$^{-1}$ and are assigned to the in-plane E$_{2g}^1$ and out-of-plane A$_{1g}$ phonon modes (inset Fig 3a), respectively. Previous Raman studies suggest that interlayer coupling softens the E$_{2g}^1$ mode through an enhanced dielectric screening dependent on the layer number, while interlayer coupling stiffens the A$_{1g}$ mode[32, 33]. Indeed, for bilayers, the E$_{2g}^1$ peak shows a constant redshift, whereas the A$_{1g}$ peak exhibits a twist angle dependent blueshift. Therefore, the separation between these two peaks ($\omega_A - \omega_E$) characterizes the effective interlayer mechanical coupling strength: the further the separation, the stronger the coupling strength. Fig. 3b shows the Raman peak separation in 44 MoS$_2$ bilayers with respect to the twist angle. Interestingly, the mechanical coupling shows a pattern very similar to that of the electronic coupling as in Fig. 2b: AA or AB stacking has the strongest coupling and the other twist angles have weaker but constant coupling.

In order to understand the unusual evolution of interlayer coupling in twisted MoS$_2$ bilayers, we perform *ab initio* calculations on the structural energy and band structures of MoS$_2$ bilayers from density functional theory within the local density approximation (LDA). A supercell approach is employed to describe the twisted bilayers, with the crystalline alignment of the top layer oriented at a commensurate angle with respect to the bottom layer. $\theta = 0°$, 60° and four intermediate commensurate angles ($\theta \sim 13°, 22°, 38°, 47°$) are studied in our calculations (Fig.

4a). When $\theta = 0°$ (AA) or $60°$ (AB), there are three high-symmetry stacking configurations: the S atoms in the top layer are placed above (1) the Mo atoms ($AA_1$ and $AB_1$), (2) the hexagonal centers ($AA_2$ and $AB_2$), or (3) the S atoms in the bottom layer ($AA_3$ and $AB_3$). As the $AA_1$ and the $AA_2$ configuration are related by spatial inversion, we use $AA_1$ to denote these two equivalent structures. Among these high-symmetry configurations, $AA_1$ and $AB_1$ have the lowest energy and are presumably the most stable (See supplementary S2 for more details). For the four twisted configurations, there are no high-symmetry configurations, and the atoms in the two layers have nearly random relative distributions. Our calculated Kohn-Sham K-valley direct bandgaps (i.e., bandgaps responsible for the direct optical transition originating from the K valley of the first Brillouin Zone of a monolayer $MoS_2$ unit cell) and the indirect bandgaps (i.e., the smallest bandgap between the valence band top and the conduction band minimum) for $AA_1$, $AB_1$, and the four twisted configurations are shown in Fig. 4b. We find that the trend in bandgap variation with the twist angle matches well with the experimental observation in Fig. 2b: The K-valley direct bandgap remains largely unchanged, but the indirect bandgap is much smaller in AA- and AB-stacked bilayers compared with all other twist angles. We note that the Kohn-Sham bandgaps should not be directly compared to the measured optical bandgaps, because of the neglect of the quasiparticle self-energy and electron-hole interaction effect[17]; however, these effects tend to cancel each other in the transition energies. The trends in the twist angle dependence should be correct.

We further examine the evolution of the calculated interlayer distance (defined by the averaged Mo-Mo interlayer separation). A strong variation of the interlayer distance on the stacking configuration is found, as illustrated in an exaggerated fashion in Fig. 4a (See detailed value in supplementary S3). The higher energy $AA_3$ and $AB_3$ configurations have the largest interlayer distance of ~ 0.68 nm; the two energetically favorable $AA_1$, and $AB_1$ configurations are found to have the smallest interlayer distance of ~0.61 nm; and all the twisted configurations have almost identical interlayer distances at ~ 0.65 nm. If we only focus on the lower energy configurations ($AA_1$, $AB_1$, and the four twisted configurations), this trend in interlayer distance evolution with twist angle is similar to the evolution of the indirect bandgap (Fig. 4b).

The configuration-dependent interlayer distance in bilayer $MoS_2$ can be understood physically by steric effects, which arise from the fact that each atom occupies a certain amount of space that strongly repulse other atoms due to a significant energy cost from overlapping electron clouds. Between two vertically stacked $MoS_2$ layers, the only interlayer adjacent atoms are S atoms. In the energetically unfavorable $AA_3$ or $AB_3$ stacking configurations, the S atoms of the top layer sit directly on the S atoms of the bottom layer in an eclipsed fashion, leading to a strong repulsion and the largest interlayer distance. In the energetically favorable $AA_1$ or $AB_1$ stacking configurations, on the other hand, the S atoms of the top layer sit on the trigonal vacancies of the S atoms of the bottom layer in a staggered fashion, resulting in reduced repulsion and the smallest interlayer distance. For other twist angles, the S atoms of the top layer sit nearly randomly relative to the S atoms of the bottom layer, and therefore, the interlayer distance is a constant and roughly at the average of that of the most stable $AA_1$ ($AB_1$) and unstable $AA_3$ ($AB_3$) stacking configurations. Consequently, the variation in interlayer distance has a simple geometric origin, which is determined by the lateral registration of adjacent S layers. Such steric effects determine the interlayer distance, which in turn modifies both the interlayer electronic and mechanical coupling strengths. This explains why these two types of coupling share similar evolutionary trends with twist angle, and they are almost linearly related (Fig. 3c).

Interestingly, if we examine separately the explicit dependence of the electronic coupling (and the corresponding indirect bandgap energy) on the interlayer vertical separation and horizontal registration, only the former has a significant effect. For example, we can artificially vary the interlayer distance of the $AA_1$-stacked bilayer in our *ab initio* calculation, and calculate the resulting changes in the direct and indirect Kohn-Sham bandgaps (dashed curves in Fig. 4c). Surprisingly, the bandgap *vs.* interlayer distance relation for different stacking configurations (symbols in Fig. 4c) all lie on the dashed curves, which are from the $AA_1$ configuration at different separation. It shows that in $MoS_2$ bilayers, the interlayer coupling strength of the electronic states near the band edges depends only explicitly on the interlayer distance, but not on the relative horizontal lattice alignment. (We note that there is an implicit dependence because the interlayer separation varies with the horizontal lattice registration). We notice that the PL peak II can vary significantly for AA and AB stacked bilayers. Presumably it arises from uncontrollable interlayer distance variation in the CVD grown bilayers. For example, the thermal expansion coefficient of $MoS_2$ is one order of magnitude larger than the fused silica substrate, which can induce a mechanical strain in the bilayer and lead to relative slipping between the two layers. This will result in a change in average distance for AA and AB stacked bilayers and variations of the indirect bandgap, but has little effect on bilayers with twisting angle between 0 and 60°.

Our discovery indicates that the steric repulsion effects, which have been extensively studied in surface reactions and nano-bio interfaces, also play an important role in understanding the electronic and vibrational properties of vdW-coupled 2D atomic layers. Such steric effects have been largely overlooked in previous studies of graphene systems[1-6]. Our calculations show that the steric effects in 2D atomic layers strongly depend on the atomic size and the in-plane atom-atom distance. Comparing $MoS_2$ with graphene, not only is the size of S atoms bigger than the size of C atoms, but also the in-plane S-S distance (0.32 nm) is much larger than the in-plane C-C distance (0.14 nm). As a result, the steric effect is about 3 times stronger in $MoS_2$ bilayers than graphene bilayers. Still, steric effects are present in graphene bilayers, where the interlayer distance of Bernal-stacked graphene bilayers is 0.01 nm smaller than that of other twisted bilayers (See supplementary S4 for more details).

In conclusion, we investigate the interlayer coupling evolution in twisted $MoS_2$ bilayers and find that the change in both electronic and mechanical interlayer coupling is relatively constant at twist angles between 0 and 60°, and becomes strongest for twist angle at 0° (AA stacking) and at 60° (AB stacking). In particular, we identify the importance of the varying interlayer distance in vdW-coupled 2D atomic layered materials. The interlayer distance can potentially be tuned through material design or external pressure. This tunability may expand the range of optical, mechanical and electrical properties of these 2D materials for device applications.

**Methods:**

**Growth of twisted MoS$_2$ bilayers**

We grew MoS$_2$ bilayers by CVD method on mica, fused silica or SiO$_2$/Si substrate using MoO$_3$ and S powder as precursors. The substrate was loaded into a 1-inch CVD furnace and placed above a crucible containing 20 mg of MoO$_3$ (≥99.5%, sigma-Aldrich). Another crucible containing 7 mg of S (≥99.5%, sigma-Aldrich) was placed 12 cm upstream from the sample. The CVD process was performed under ambient pressure while flowing ultrahigh-purity nitrogen. The recipe is: sit at 105 ℃ with 500 sccm for one hour, ramp to 700 ℃ at 15 ℃/min with 10~15 sccm N$_2$, sit at 700 ℃ for 5~10 mins, and then cool down naturally with 500 sccm gas flow. The ratio of MoS$_2$ bilayer to monolayer increases with the growth time. Bilayers with yield as high as 30% can be achieved when performing 10 min growth under 700 ℃. In general, the yield of AA-stacked bilayer of $\theta = 0°$ is the highest (~85%), followed by AB-stacked bilayer of $\theta = 60°$ (~10%).

**Sample Characterization**

(a) Difference reflection spectra: A supercontinuum laser (470 nm ~ 1800 nm) was used as the light source. We focused the supercontinuum to a diameter of ~ 2 μm on the sample, and analyzed the reflected light with a spectrograph equipped with an array charged coupled device (CCD) detector. Two sets of reflection spectra with the MoS$_2$ monolayers or bilayers inside the laser beam ($I_{\text{inside}}$) and outside the beam ($I_{\text{outside}}$) were taken and the final difference spectrum was obtained as $\Delta R/R = (I_{\text{inside}} - I_{\text{outside}})/I_{\text{outside}}$.

**(b)** SHG measurements: A femtosecond laser (ATSEVA TiF-100F laser, 80 MHz repetition rate, wavelength at 800 nm, pulse duration of 100 fs) was used as the light source. We used 20 mW linearly polarized light focused to a diameter of ~ 5 $\mu m$ for excitation, and collected the SHG signal using a spectrometer coupled with a CCD.

**(c)** Photoluminescence and Raman spectra were taken by a Horiba HR800 system with laser excitation wavelength of 532 nm. The laser is focused to a diameter of ~ 1 μm and has a power of ~ 1 mW.

*Ab initio* **calculations**

*Ab initio* calculations of twisted $MoS_2$ bilayers were performed using density functional theory (DFT) in the local density approximation (LDA) implemented in the Quantum Espresso package[34]. A supercell arrangement was used, with the cell dimension in the out of plane direction set at 20 Å to avoid interactions between the $MoS_2$ bilayer and its periodic images. We employ norm-conserving pseudopotentials, with a plane-wave energy cutoff of 140 Ry. The structures were fully relaxed until the force on each atom is smaller than 0.01 eV/Å. Spin-orbit coupling was not included in our calculations.

We also carried out the theoretical calculations for different bilayer configurations using the PBE-D exchange-correlation functional. The interlayer distances obtained with the PBE-D method are almost identical to that from LDA calculations (See supplementary S3 for more details). Consequently, the evolution of bandgaps calculated using the PBE-D method shows a similar trend as that obtained from LDA calculations (See supplementary S5 for more details).

**Acknowledgement**: This study was supported by Office of Basic Energy Science, Department of Energy under contract No. DE-SC0003949 (Early Career Award) and No. DE-AC02-05CH11231 (Materials Science Division). Research supported in part by the Theory Program at Lawrence Berkeley National Lab through the Office of Basic Energy Sciences, U.S. Department of Energy (DOE) under Contract No. DE-AC02-05CH11231 which provided code developments and simulations, and by the National Science Foundation under Grant No. DMR10-1006184 which provided structural study and analysis of interlayer coupling. SGL acknowledges support of a Simons Foundation Fellowship in Theoretical Physics. Computation resources at NERSC funded by DOE are used. K. Liu acknowledges support from National Program for Thousand Young Talents of China.

**Figure captions:**

**Figure 1 | Twisted MoS$_2$ bilayers. a-d**, Schematics of the atomic structure of a MoS$_2$ monolayer (a) and bilayers with twist angles of $\theta = 0°$ (b), $\theta = 15°$ (c), and $\theta = 60°$ (d). **e-h**, Optical reflection images of a MoS$_2$ monolayer and twisted bilayers corresponding to (a-d). **i**, Difference reflection spectra ($\Delta R/R$) from the monolayer and bilayers corresponding to (e-h). The two resonance peaks are characteristics of direct exciton transitions in MoS$_2$. **j**, Second-harmonic generation (SHG) signal from a MoS$_2$ monolayer, an AA-stacked bilayer ($\theta = 0°$) and an AB-stacked bilayer ($\theta = 60°$). The traces in (i-j) are shifted vertically for clarification.

**Figure 2 | Probing interlayer electronic coupling in twisted MoS$_2$ bilayers. a**, Photoluminescence (PL) spectra of MoS$_2$ monolayer and bilayers with twist angles of $\theta = 0°, 15°, 60°$. **b**, Dependence of PL peak energies on the twist angle for 44 MoS$_2$ bilayers. The peak I energy is almost identical for all bilayers. The peak II energy is lowest for AA- and AB-stacked bilayers, and is higher but at a nearly constant value for other twist angles. **c,d** Calculated local density approximation (LDA) Kohn-Sham band structure of a MoS$_2$ monolayer (c) and of the most energetically favorable AA-stacked bilayer (d). Transition I is associated with the K-valley direct bandgap (peak I in PL spectra). The indirect bandgap transitions II and II' have nearly degenerate energy and either may be associated with the peak II in the PL spectra.

**Figure 3 | Probing interlayer mechanical coupling in twisted MoS$_2$ bilayer. a**, Raman spectra of a MoS$_2$ monolayer and bilayers with twist angles of $\theta = 0°, 15°, 60°$. **b**, Raman peak separation between the A$_{1g}$ and E$_{2g}^1$ ($\omega_A - \omega_E$) in MoS$_2$ bilays of different twist angles, which is the largest for AA or AB stacking and nearly a constant for other twist angles. **c,** The relation between peak II in Fig. 2b and Raman peak separation. The data collapse onto a linear line, indicating an intrinsic correlation between electronic and mechanical coupling.

**Figure 4 | Structural and bandgap calculations of MoS$_2$ bilayers. a**, Schematics of MoS$_2$ bilayers with AA, AB, and different twisted configurations. Mo atoms are shown as green spheres; two S atoms of the same horizontal position are presented by one yellow sphere. Interlayer distance variations are exaggerated for illustration. The interlayer separation is ~ 0.61 nm for AA$_1$, AB$_1$ stacking, ~ 0.62 nm for AB$_2$ stacking, ~ 0.68 nm for AA$_3$ and AB$_3$ stacking, and ~ 0.65 nm for the four twisted bilayers. **b**, Calculated values for the Kohn-Sham K-valley direct bandgap (orange) and indirect bandgap (dark yellow) for the energetically favorable structures at each twist angle. **c**, The calculated Kohn-Sham bandgap as a function of the interlayer distance for different stacking configurations (symbols). The dashed curve shows the bandgap values for AA$_1$ stacking with artificially varied interlayer separation. Their comparison indicates that the electronic coupling in MoS$_2$ bilayer is largely determined by the interlayer separation, and does not depend explicitly on the horizontal registration of the two layers.

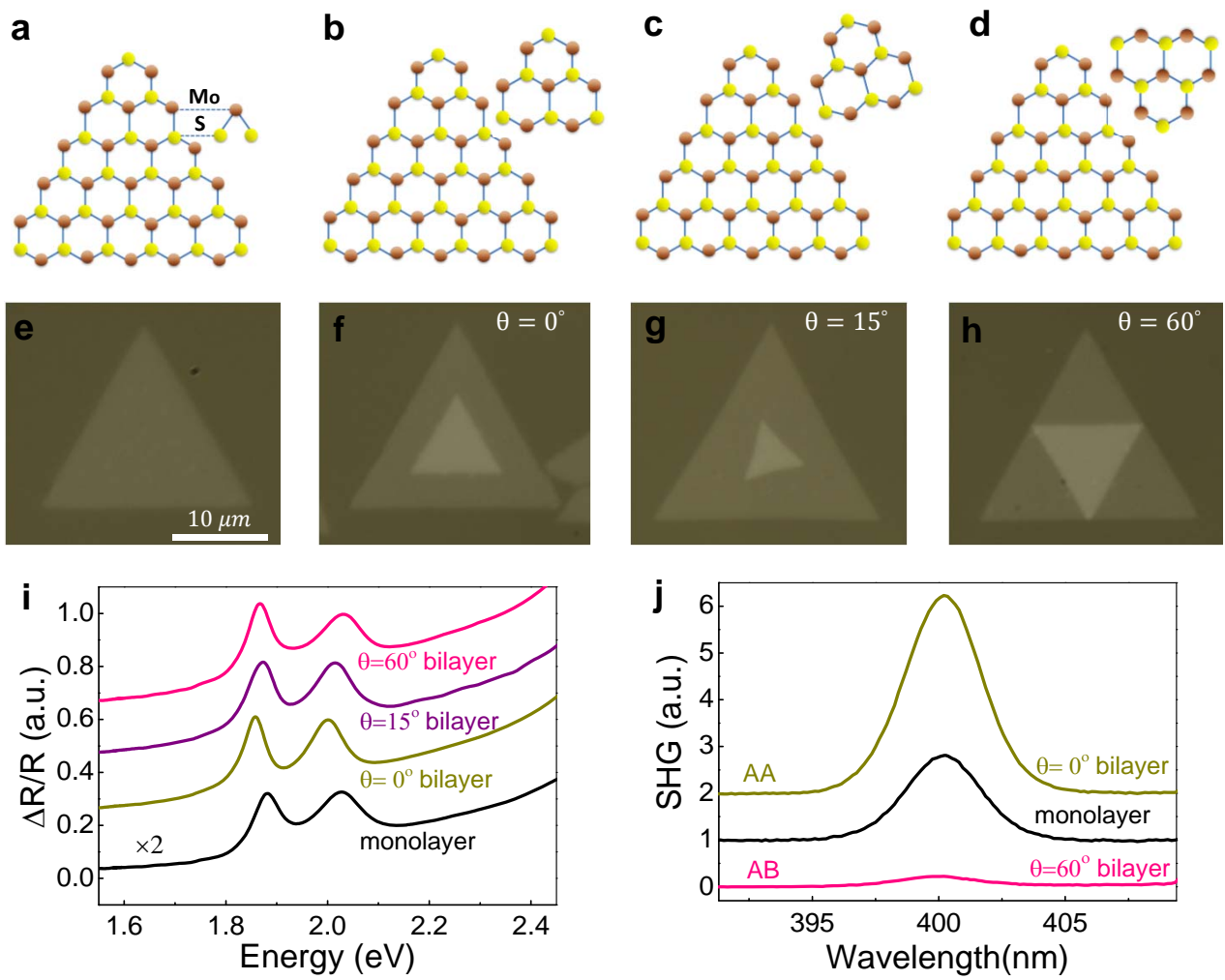

**Figure 1**

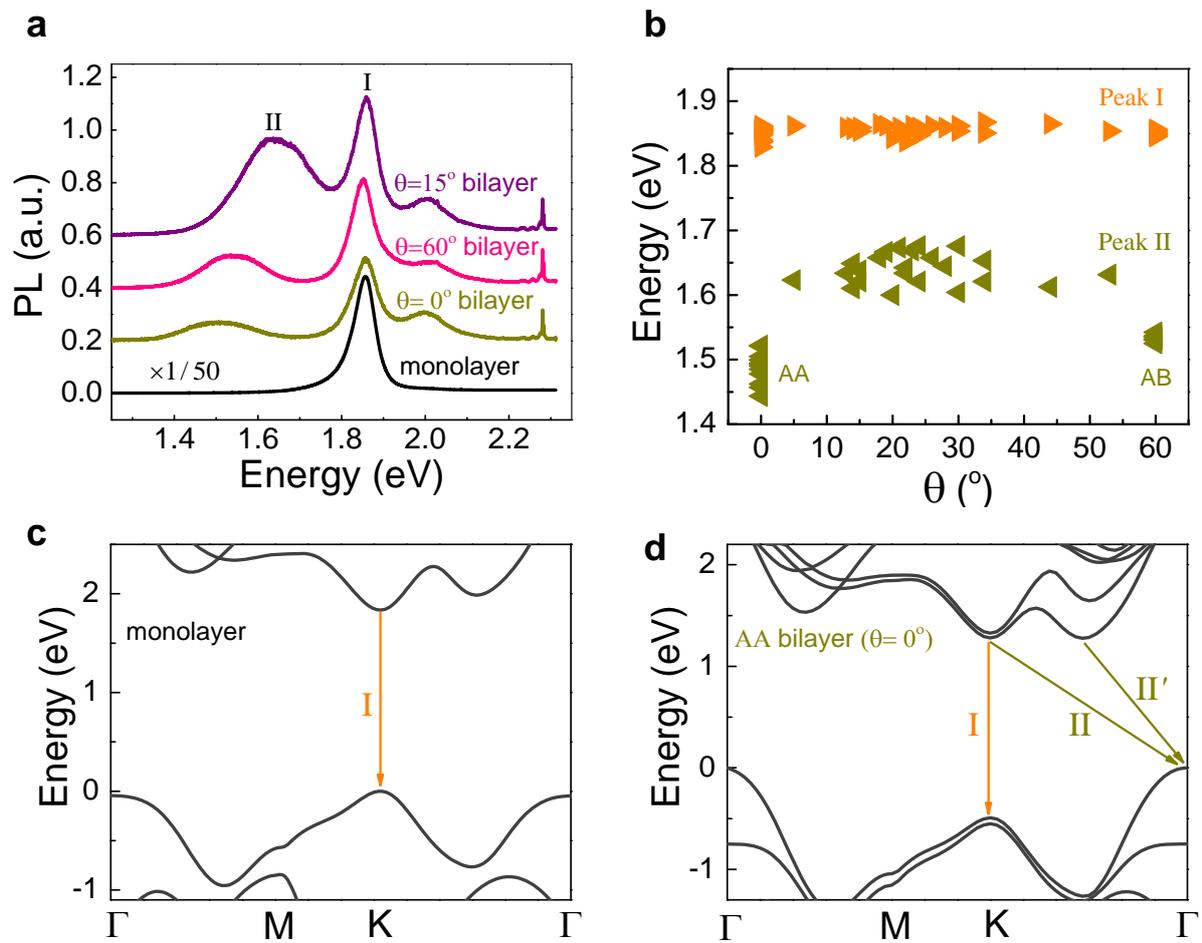

**Figure 2**

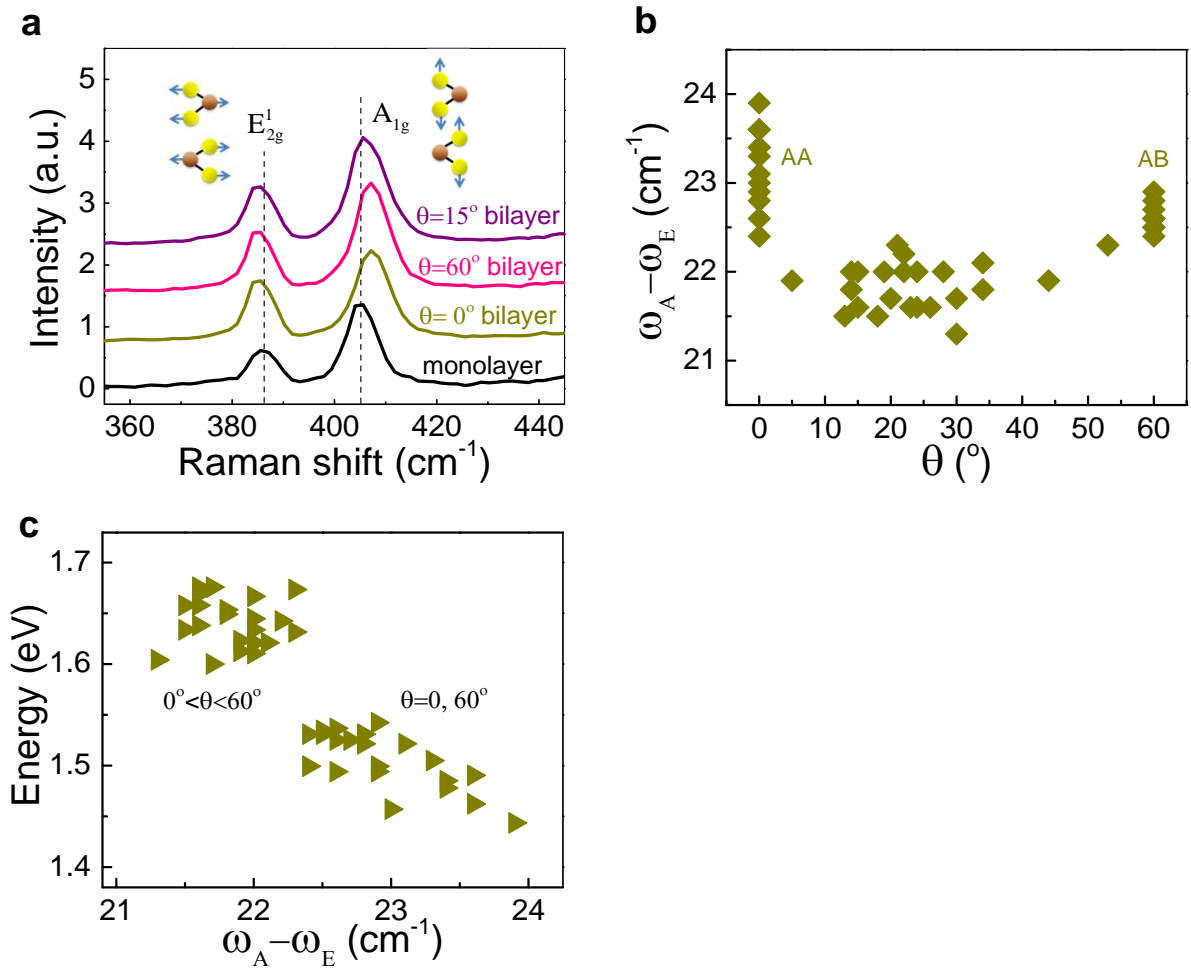

Figure 3

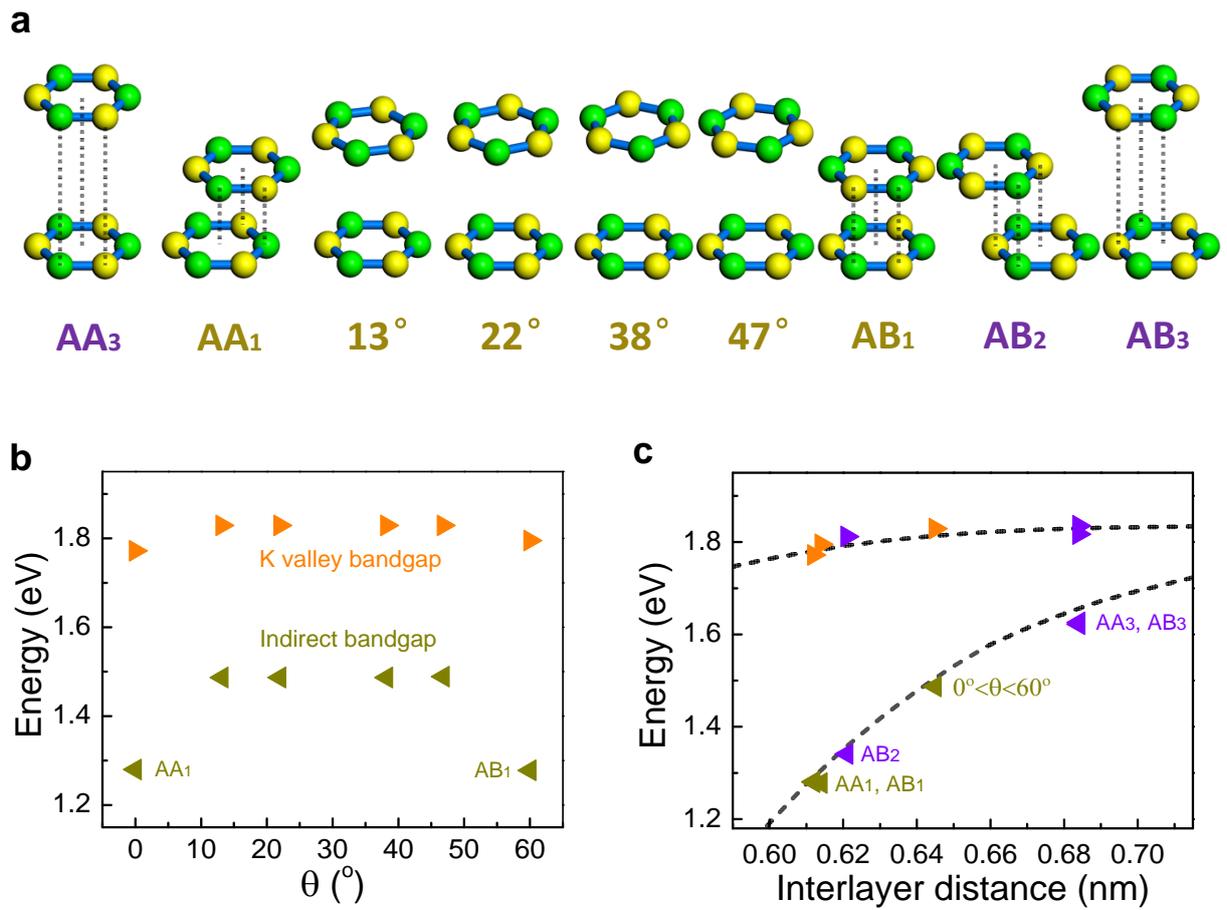

**Figure 4**

Supplementary information for

# Evolution of Interlayer Coupling in Twisted MoS$_2$ Bilayers

Kaihui Liu[1,2]*, Liming Zhang[3]*, Ting Cao[1,4]*, Chenhao Jin[1], Diana Qiu[1,4], Qin Zhou[4],

Alex Zettl[1,4,5], Peidong Yang[3,5,6], Steve G. Louie[1,4] and Feng Wang[1,4,5]

**S1. Determination of stacking order for $\theta = 0°, 60°$ bilayers by SHG**

**S2. LDA energy stability of different configurations in twisted MoS$_2$ bilayers**

**S3. Calculated interlayer distance in twisted MoS$_2$ bilayers**

**S4. Calculated interlayer distance in twisted graphene bilayers**

**S5. PBE-D band gap evolution of twisted MoS$_2$ bilayers**

## S1. Determination of stacking order for $\theta = 0°, 60°$ bilayers by SHG

The bilayers with twist angle of $\theta = 0°, 60°$ can have AA or AB stacking. In the AA stacking case, the in-plane Mo-S bond direction is the same for the two layers, and in the AB-stacking case the in-plane Mo-S bond direction is opposite for the two layers. In order to distinguish these two stacking orders, we employ nonlinear optical second-harmonic generation (SHG), which is very sensitive to asymmetry of the surface layers[1]. Monolayer $MoS_2$ belongs to the noncentrosymmetric $D_{3h}^1$ group and therefore has a strong SHG signal (black curve in Fig. 1j). In contrast, AB-stacked bilayer $MoS_2$ belongs to the centrosymmetric $D_{3d}^1$ group, and therefore, has no appreciable SHG signal (pink curve in Fig. 1j, $\theta = 60°$). Owing to the absence of an inversion center, the SHG signal is very strong for AA-stacked bilayer $MoS_2$ (dark orange curve in Fig. 1j, $\theta = 0°$), roughly two times stronger than for the monolayer. The fact that the $\theta = 0°$ ($\theta = 60°$) bilayer is AA (AB) stacking confirms that the two layers have the same edge types[2-4].

## S2. LDA energy stability of different configurations in twisted $MoS_2$ bilayers

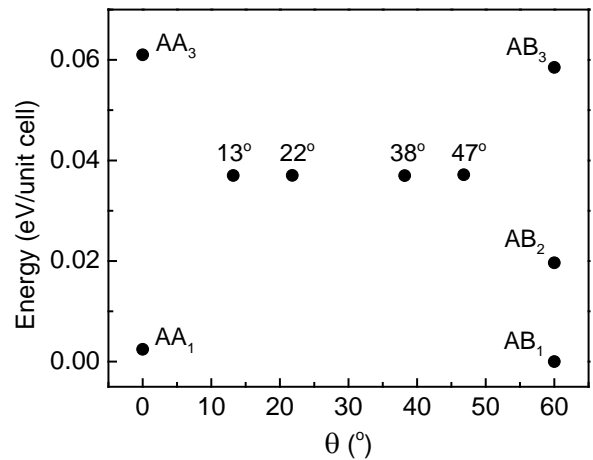

Figure S1. Relative energy of different configurations as calibrated to the lowest energy $AB_1$ configuration.

Figure S1 shows the relative energy stability of each configuration at the LDA level. Consistent with the stacking order of a natural $MoS_2$ crystal, we find that the $AB_1$ configuration has the lowest energy. $AA_1$ has nearly degenerate energy, which is slightly higher than the $AB_1$ configuration. It is worth noting that all twist structures are almost degenerate in energy.

### S3. Calculated interlayer distance in twisted $MoS_2$ bilayers

We used both LDA and the PBE-D exchange-correlation functionals, with the latter explicitly including the effects of long-range van der Waals interaction, to calculate the interlayer distances of different configurations[5]. Our results are summarized in Table S1. The two methods gave very similar results with interlayer distances differing by less than 0.1 Å.

Table S1 Interlayer distance of $MoS_2$ bilayer in different configurations

| Configuration | LDA interlayer distance (Å) | Van der Waals functional (PBE-D) interlayer distance (Å) |
|---|---|---|
| $AA_1$ | 6.1 | 6.2 |
| $AA_3$ | 6.8 | 6.8 |
| $AB_1$ | 6.1 | 6.2 |
| $AB_2$ | 6.2 | 6.2 |
| $AB_3$ | 6.8 | 6.8 |
| 13.2° | 6.5 | 6.5 |
| 21.8° | 6.5 | 6.5 |
| 38.2° | 6.5 | 6.5 |
| 46.8° | 6.5 | 6.5 |

### S4. Calculated interlayer distance in twisted graphene bilayers

For graphene bilayer of $\theta = 0°$, there are two high-symmetric configurations: C atom of the top layer is placed on the hexagonal centre of bottom layer (Bernal stacking), or directly on the C atom of the bottom layer (higher energy). Since two bipartite sites are equivalent, the twist angle dependence in graphene bilayer can be described within half of that of the MoS$_2$ bilayer, namely 30°.

Table S2 Interlayer distance of graphene bilayer in different configurations

| Configuration | LDA interlayer distance (Å) | Van der Waals (PBE-D) interlayer distance (Å) |
| --- | --- | --- |
| 0° (higher energy) | 3.5 | 3.4 |
| 0° (Bernal stacking) | 3.3 | 3.2 |
| 13.2° | 3.4 | 3.3 |
| 21.8° | 3.4 | 3.3 |

## S5. PBE-D band gap evolution of twisted MoS$_2$ bilayers

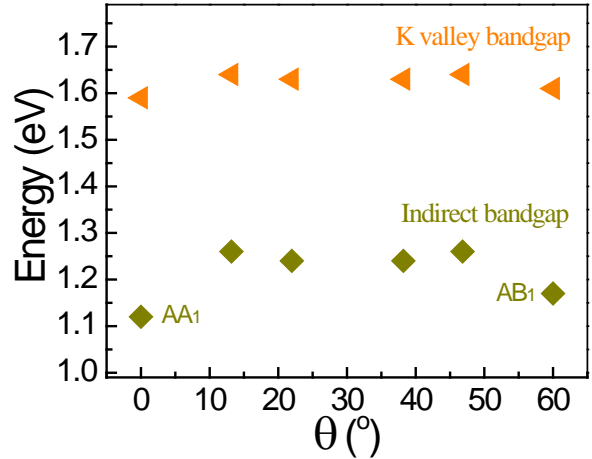

Figure S2. Calculated values for the Kohn-Sham direct bandgap at the K valley (orange) and indirect bandgap (dark yellow) for the energetically favorable structures at each twist angle.

We used the PBE-D exchange-correlation functional to calculate the K-valley direct bandgap and indirect bandgap for the energetically favorable structures at each twist angle (Fig. S2). Comparing with the LDA results, the PBE-D bandgaps are smaller. However, the trend of bandgap evolution with the twist angle in the PBE-D method is the same as that in the LDA method.